\documentstyle[aps,epsfig,floats]{revtex}
\newcommand{\be}{\begin{equation}}
\newcommand{\ee}{\end{equation}}
\newcommand{\bea}{\begin{eqnarray}}
\newcommand{\eea}{\end{eqnarray}}

\newcommand{\NP}[1]{{\it Nucl.\ Phys.}\ {\bf #1}}
\newcommand{\ZP}[1]{{\it Z.\ Phys.}\ {\bf #1}}
\newcommand{\PL}[1]{{\it Phys.\ Lett.}\ {\bf #1}}
\newcommand{\PR}[1]{{\it Phys.\ Rev.}\ {\bf #1}}
\newcommand{\PRL}[1]{{\it Phys.\ Rev.\ Lett.}\ {\bf #1}}

\begin{document}
\tighten
\thispagestyle{empty}
\title{
\begin{flushright}
\begin{minipage}{4 cm}
\small
DCPT/01/12\\
DFTT 1/2001\\
INFNCA-TH0101\\
IPPP/01/06\\
\end{minipage}
\end{flushright}
\vspace{5mm}
Parton Densities and Fragmentation Functions \\ from Polarized 
\mbox{\boldmath $\Lambda$}
Production in Semi-Inclusive DIS
\protect} 
\vspace{5mm}
\author{M. Anselmino$^1$, M. Boglione$^2$, U. D'Alesio$^3$, E. 
Leader$^{4,}\!\!$
\footnote{Permanent address: High Energy Physics Group, Blackett 
Laboratory, 
Imperial College,\\ \mbox{} \hspace*{3.15cm}Prince Consort Road, London 
SW7 2BW, 
United Kingdom}, F. Murgia$^3$
\vspace{5mm}
\mbox{}\\
{\it $^1$ Dipartimento di Fisica Teorica, Universit\`a di Torino and \\
          INFN, Sezione di Torino, Via P. Giuria 1, I-10125 Torino, 
Italy}\\
\vspace{0.3cm}
{\it $^2$ Department of Physics, University of Durham, Science 
Laboratories,}\\
{\it South Road, Durham DH1 3LE, United Kingdom}\\
\vspace{0.3cm}
{\it $^3$ INFN, Sezione di Cagliari and Dipartimento di Fisica,  
Universit\`a di Cagliari,\\
C.P. 170, I-09042 Monserrato (CA), Italy}\\
\vspace{0.3cm}
{\it $^4$ Physics Department, Theory Division, Vrije Universiteit 
Amsterdam,}\\
{\it De Boelelaan 1081, 1081 HV Amsterdam, The Netherlands}}

\maketitle

\vspace{1cm}

\begin{abstract}
We consider the longitudinal polarization of $\Lambda$
and  $\bar \Lambda$ produced in the current fragmentation region
of polarized deep inelastic scattering. We show how the various
cross sections can be used to test the 
underlying parton dynamics, and how one can 
extract information about certain parton densities which are poorly 
known, in particular the polarized strange density sum
$\Delta s(x)+\Delta\bar s(x)$, and about fragmentation functions
which are totally unknown and which are 
difficult to access by other means. We show also how one can obtain
information concerning the intriguing question as to whether
$s(x)=\bar s(x)$ and whether $\Delta s(x)=\Delta\bar s(x)$.
\end{abstract}

\vspace{0.6cm}

~~~~~~~~~~PACS numbers: 13.60.Hb, 13.85.Ni, 13.87.Fh, 13.88.+e

\section{Introduction}

As has been emphasized in several papers
\cite{abm00,jaf,ekk,deflo2,kbv,kot,bel,blt,al,mssy}, measurements of the 
polarization of $\Lambda$ baryons produced in high energy deep 
inelastic lepton-hadron collisions offer an excellent test of the 
dynamics of spin transfer from partons to hadrons.

In this paper we consider all possible semi-inclusive reactions 
involving unpolarized or longitudinally polarized leptons and nucleons, 
with or without the measurement of the longitudinal polarization of  
$\Lambda$ or $\bar \Lambda$ produced in the current fragmentation 
region. We draw the attention to another important aspect of such 
reactions, namely the information obtainable about the polarized parton 
densities, and about the unpolarized and polarized fragmentation 
functions $D_q^{\Lambda}$, $\Delta D_q^{\Lambda}$ for a quark into a  
$\Lambda$ (or $\bar \Lambda$).
Despite recent progress \cite{y} some polarized parton densities are 
still relatively poorly determined.
In principle, in semi-inclusive DIS, one can obtain information about 
the polarized strange quark density $\Delta s (x) + \Delta \bar s (x)$, 
which is poorly known, and also about the interesting question as to 
whether $s(x) \neq \bar s(x)$ and $\Delta s(x) \neq \Delta \bar s(x)$.

Regarding the fragmentation functions $D_q^{\Lambda}$ and 
$\Delta D_q^{\Lambda}$ very little is known. 
Indeed the $\Delta D_q^{\Lambda}$ are not constrained at all by the 
present 
$e^+e^-$ data.
These can, in principle, all be determined in semi-inclusive DIS. They 
could 
also be accessed in the reaction $p p \to \Lambda X$ with a polarized 
proton 
beam or target \cite{deflo1}.

In this paper, as in \cite{abm00}, we work in LO QCD. Given the 
preliminary state of experiments in this field it does not seem 
sensible at this point to undertake the extremely complicated NLO 
analysis \cite{deflo2} or the somewhat simpler version given in 
\cite{x}. However, as emphasized in \cite{x}, it is important to remain 
vigilant about inaccuracies caused by using the LO formalism, and great 
attention should be paid to the various tests of the reliability of the 
LO treatment given in the following. For a more general discussion of 
this question and suggestions concerning the estimation of theoretical 
errors generated by the LO treatment, see Ref.\cite{x}.

To really extract the maximum of information from these reactions one 
should try to study the triply-differential cross section 
$d\sigma/dxdydz$ where $x$, $y$, $z$ are the usual semi-inclusive DIS 
variables \cite{abm00}. In this it is the $y$-dependence that  tests 
the dynamics, whereas the parton densities and fragmentation functions 
emerge from the $x$ and $z$-dependence. 
 
In Section II we define precisely what cross sections and polarizations 
we wish to consider. As mentioned we deal only with longitudinal 
(helicity) polarization of the leptons, nucleons and $\Lambda$'s.

In Section III we introduce modified differential cross sections which 
are simply related to parton model soft functions and which allow tests 
of the underlying parton dynamics.

In Section IV we study in detail what information can be extracted about 
the  parton densities and fragmentation functions.
Conclusions follow in Section V.

\section{The independent observables of the reaction}

We consider the reaction 
\be
\ell(\lambda) + N(\mu) \to H(h) + \ell' + X
\label{reaction}
\ee 
of a charged lepton $\ell$ with helicity $\lambda = \pm 1/2$ on a 
nucleon $N$ 
of helicity $\mu = \pm 1/2$ producing, semi-inclusively, a spin $1/2$ 
hyperon 
$H$ with helicity $h=\pm 1/2$ . The hyperon H is such that its 
polarization 
can be determined from its decay distribution.
We consider kinematical regions where $Z$ exchange is negligible.

The fundamental invariant differential cross section will be written as
\be
\frac{d\sigma _{\lambda \mu} ^{H_h}}{dx\,dy\,dz}\,,
\label{fund-cross}
\ee
where $x$, $y$, $z$ are the usual semi-inclusive DIS variables. 
In the following we do not include the differentials $dx\,dy\,dz$ 
unless 
necessary for clarity.

The cross section for simply producing $H$ from a given initial 
helicity 
state is given by
\be
d\sigma _{\lambda \mu} ^{H} = d\sigma _{\lambda \mu} ^{H_+} + 
                              d\sigma _{\lambda \mu} ^{H_-}\,.
\ee
The longitudinal or helicity polarization of H, as produced from a 
given 
initial state $(\lambda \mu)$ is then given by 
\be
P_{\lambda \mu}^{H} = 
\frac{d\sigma _{\lambda \mu}^{H_+} -  d\sigma _{\lambda \mu}^{H_-}}
{d\sigma _{\lambda \mu}^{H}}\,.
\ee

Parity invariance of the strong and electromagnetic interactions 
implies that 
\be
d\sigma _{\lambda \mu} ^{H_h} = d\sigma _{-\lambda -\mu} ^{H_{-h}}\,,
\label{parity1}
\ee
so that
\be
P_{-\lambda -\mu}^{H} = - P_{\lambda \mu}^{H}\,.
\label{parity2}
\ee
Of course for an {\it unpolarized} initial state in a parity conserving 
theory, the longitudinal polarization must vanish, as is clear from 
(\ref{parity2}).
Thus $P^H$ really measures the spin transfer from either lepton or 
nucleon to 
$H$.

Cross sections or polarizations relevant to an unpolarized lepton 
and/or an unpolarized nucleon are indicated by a zero. For example, for 
an unpolarized lepton we have
\be
d\sigma _{0 \mu}^{H_h} = \frac{1}{2} (\,d\sigma _{+ \mu}^{H_h} + 
d\sigma _{- \mu}^{H_h}\,)\,.
\ee
Because of (\ref{parity1}) there will be only four independent cross 
sections or observables, instead of the original $2 \times 2 \times 2 = 
8$.

We shall take as the four independent cross sections:
\begin{description}
\item[a)]
The unpolarized cross section
\be
d\sigma ^H \equiv d\sigma _{00}^H = \frac{1}{4} (\,d\sigma _{++}^H + 
d\sigma _{--}^H + d\sigma _{+-}^H + d\sigma _{-+}^H\,) = 
\frac{1}{2} (\,d\sigma _{++}^H + d\sigma _{+-}^H\,) \,.
\label{a}
\ee
\item[b)]
The target-spin dependent cross section difference
\be
\Delta d \sigma ^H \equiv d\sigma _{++}^{H} - d\sigma _{+-}^H \,.
\label{b}
\ee
\item[c)]
The spin-transfer cross section from a polarized lepton with an 
unpolarized nucleon 
\be
d\sigma _{+0}^{H_+\,-\,H_-} \equiv d\sigma _{+0}^{H_+} -  d\sigma 
_{+0}^{H_-} =
P_{+0}^{H}\,d\sigma^{H}
\,,
\label{c}
\ee
\noindent
where we have chosen a positive helicity for the lepton
and have used, via (\ref{parity1}),
\be
d\sigma_{+0}^{H} = d\sigma_{-0}^{H} =
d\sigma_{0+}^{H} = d\sigma_{0-}^{H} = d\sigma^{H}\,.
\label{c2}
\ee
\noindent
Clearly, also via  (\ref{parity1})
\be
d\sigma _{+0}^{H_+ \,- H_-} = - d\sigma _{-0}^{H_+ \,- H_-}\,.
\ee
\item[d)]
The spin-transfer cross section from a polarized nucleon with an 
unpolarized lepton
\be
d\sigma _{0+}^{H_+ \,- H_-} \equiv  d\sigma _{0+}^{H_+} - d\sigma 
_{0+}^{H_-}
= P_{0+}^{H}\,d\sigma^{H}
\,.
\label{d}
\ee
\end{description}

Since these are four linearly independent observables, all others can be 
written in terms of them. For example it is easy to see that 
\be
d\sigma _{++}^{H_+ \,- H_-} = d\sigma _{+0}^{H_+ \,- H_-} + d\sigma 
_{0+}^{H_+ \,- H_-}
\label{s1}
\ee
and 
\be
d\sigma _{+-}^{H_+ \,- H_-} = d\sigma _{+0}^{H_+ \,- H_-} -
d\sigma _{0+}^{H_+ \,- H_-}\,.
\label{s2}
\ee
In terms of the hyperon polarization, (\ref{s1}) and (\ref{s2}) imply 
\be
P_{++}^{H} = \frac{1}{2} 
\left(\, 1+\frac{d\sigma_{+-}^H }{d\sigma_{++}^H} \,\right)
\left(\, P_{+0}^{H} + P_{0+}^{H}\, \right) 
\label{P++}
\ee
and
\be
P_{+-}^{H} = \frac{1}{2}
\left(\, 1+\frac{d\sigma_{++}^H}{d\sigma_{+-}^H}\, \right)
\left(\, P_{+0}^{H} -  P_{0+}^{H} \,\right) \,.
\label{P+-}
\ee

Note that equations like (\ref{P++}) and (\ref{P+-}) are not
predictions of the detailed dynamics, but follow from parity
invariance of the strong and electromagnetic interactions.
Similar relations were given in \cite{abm00}.

\section{ The Dynamical Model}

In LO pQCD, the general cross section (\ref{fund-cross}) corresponding 
to the process in (\ref{reaction}) is given by 
\be
\frac{d\sigma_{\lambda \mu}^{H_h}}{dx \,dy \,dz} = 
\sum _{q,\lambda _q} e^2_q \,q^{\mu}_{\lambda _q}(x)\,
\frac{d\hat\sigma _{\lambda \lambda_q}}{dy}\,
D_{q_{ \lambda _q}}^{H_h}(z) \,,
\label{gen-cross}
\ee
where the sum is over quarks and antiquarks, $q^{\mu}_{\lambda _q}(x)$ 
is the parton number density for quarks of helicity $\lambda _q$ in a 
proton of helicity $\mu$, while $D_{q_{ \lambda_q}}^{H_h}$ is the 
fragmentation function for a quark $q$ of helicity $\lambda _q$ to 
fragment into hyperon $H$ with helicity $h$; $d\hat\sigma _{\lambda 
\lambda_q}/dy$ is the lepton-quark (or antiquark) differential cross 
section for an initial state with the lepton having helicity $\lambda$ 
and the quark (or antiquark) helicity $\lambda_q$.
The simple helicity structure of (\ref{gen-cross}) reflects the fact 
that helicity is conserved for massless quarks in $\ell q \to \ell q$.

There are two independent partonic cross sections
\bea
\frac{d\hat\sigma _{++}}{dy} &=& 
\frac{d\hat\sigma _{--}}{dy}  =
\frac{4 \pi \alpha ^2}{sxy^2} \,,
\\
\frac{d\hat\sigma _{+-}}{dy} &=& 
\frac{d\hat\sigma _{-+}}{dy} =
\frac{4 \pi \alpha ^2}{sxy^2}(1-y)^2 \,,
\eea
where $s$ is the squared centre of mass energy corresponding to the 
process in 
Eq. (\ref{reaction}). 

In order to simplify the expressions needed for the flavour analysis, 
we renormalize our four independent cross sections (\ref{a}-\ref{c}),
(\ref{d}), by dividing out certain common kinematic factors.
Thus we work with

\begin{description}
\item[a$^{\prime}$)] \mbox{}\be
d\tilde\sigma ^H (x,z) \equiv 
\Big[\frac{2\pi \alpha ^2}{sx} 
\,\frac{1+(1-y)^2}{y^2}\Big]^{-1}d\sigma^H = 
\sum _q e^2_q  \,q(x) \, D_q^H(z)\,,
\label{a'}
\ee
where $q(x)$ and $D_q^H(z)$ are the usual unpolarized parton density 
and fragmentation functions respectively, $q(x) = q^+_+(x) + 
q^+_-(x)$, $D_q^H(z) = D_{q_+}^{H_+}(z) + D_{q_+}^{H_-}(z)$.
\item[b$^{\prime}$)] \mbox{} \be
\Delta d\tilde\sigma ^H (x,z) \equiv 
\Big[\frac{4\pi \alpha ^2}{sx} \, \frac{y(2-y)}{y^2}\Big]^{-1}
\Delta d\sigma^H = \sum _q e^2_q  \,\Delta q(x) \, D_q^H(z)\,,
\label{b'}
\ee
where $\Delta q(x)=q^+_+(x)-q^+_-(x)$ is the usual longitudinally
polarized parton density.
\item[c$^{\prime}$)] \mbox{} \be
d\tilde\sigma_{+0}^{H_+ - H_-} (x,z) \equiv 
\Big[\frac{2\pi \alpha ^2}{sx} \, \frac{y(2-y)}{y^2}\Big]^{-1}
d\sigma_{+0}^{H_+ - H_-}= 
\sum _q e^2_q \, q(x)  \,\Delta D_q^H(z)\,,
\label{c'}
\ee
where 
\be
\Delta D_q^H(z) = D_{q_+}^{H_+}(z) -  D_{q_+}^{H_-}(z) = 
                D_{q_+}^{H_+}(z) -  D_{q_-}^{H_+}(z)  \,.
\ee
\item[d$^{\prime}$)] \mbox{} \be
d\tilde\sigma_{0+}^{H_+ - H_-} (x,z) \equiv 
\Big[\frac{2\pi \alpha ^2}{sx} 
\frac{1+(1-y)^2}{y^2}\Big]^{-1}d\sigma_{0+}^{H_+ - H_-}= 
\sum _q e^2_q\,\Delta q(x)\,\Delta D_q^H(z)\,.
\label{d'}
\ee
\end{description}
We see that for each flavour, for a given hyperon $H$ and for
a given target, the four independent cross sections just 
correspond to different combinations of the four independent soft 
functions, $q(x)$, $\Delta q(x)$, $D^H_q(z)$, $\Delta D^H_q(z)$.

It should be noted that the fact that $d\tilde\sigma ^H$,  $\Delta 
d\tilde\sigma ^H$, $d\tilde\sigma _{+0} ^{H_+-H_-}$,
$d\tilde\sigma _{0+} ^{H_+-H_-}$ 
depend only on $x$ and $z$ (neglecting the known and mild dependence
on $Q^2=xys$, due to QCD evolution) is a direct consequence of the 
parton dynamics and should be tested experimentally.

\section{Extraction of parton densities and fragmentation 
functions}

We assume that the usual unpolarized parton densities $u(x)$, $d(x)$, 
$\bar u(x)$, $\bar d(x)$ are reasonably well known and can be used as 
{\it input} in the following expressions. 

We consider the production of $\Lambda$ and  $\bar \Lambda$ hyperons on 
both proton and neutron targets and show how one can systematically 
obtain information about the parton densities $s(x)$, $\bar s(x)$,  
$\Delta s(x)$, $\Delta \bar s(x)$ and about the fragmentation functions 
$\Delta D^{\Lambda}_u(z)$,
$\Delta D^{\Lambda}_d(z)$, $\Delta D^{\Lambda}_s(z)$. 

We shall assume good enough control over systematic errors to allow us 
to combine cross sections for different targets and for $\Lambda$ and 
$\bar \Lambda$ final particles. This is a non-trivial experimental 
issue, but well worth the effort, since it then becomes possible to 
obtain very simple expressions for the parton densities and 
fragmentation functions under study.

\subsection{Unpolarized cross section}

Using only charge conjugation invariance,
\be
D_q^{\Lambda} = D_{\bar q}^{\bar \Lambda}\,,
\ee
and isospin invariance,
\be
D_d^{\Lambda} = D_u^{\Lambda}\,,
\ee
we obtain from (\ref{a'}) the well known relations \cite{x}
\bea
d\tilde\sigma ^{\Lambda+\bar \Lambda} \Big| _p - 
d\tilde\sigma ^{\Lambda+\bar \Lambda} \Big| _n &=& 
\frac{1}{3}[u(x)+\bar u(x)-d(x)-\bar d(x)] \,D_u^{\Lambda + \bar 
\Lambda}(z)\,,
\label{d-} \\
d\tilde\sigma ^{\Lambda-\bar \Lambda} \Big| _p - 
d\tilde\sigma ^{\Lambda-\bar \Lambda} \Big| _n &=& 
\frac{1}{3}[u_v(x)-d_v(x)] \, D_u^{\Lambda - \bar \Lambda}(z) \,,
\label{d+}
\eea
where $q_v(x)$ is a valence quark density.

Measurements of the cross section differences on the LHS of (\ref{d-}) 
and (\ref{d+}) thus enable a determination of $D_u^{\Lambda + 
\bar\Lambda}$ and $D_u^{\Lambda - \bar\Lambda}$ and therefore of the 
individual  $D_u^{\Lambda}$ and $D_u^{\bar \Lambda}$.

Next we consider the combinations
\bea
d\tilde\sigma ^{\Lambda+\bar \Lambda} \Big| _p + 
d\tilde\sigma ^{\Lambda+\bar \Lambda} \Big| _n &=& 
\frac{5}{9}\,[u(x)+\bar u(x)+d(x)+\bar d(x)] \, D_u^{\Lambda + \bar 
\Lambda}(z)
\nonumber \\
&+&\frac{2}{9}\,[s(x)+\bar s(x)] \, D_s^{\Lambda+\bar \Lambda}(z)\,,
\label{s+}
\\
d\tilde\sigma ^{\Lambda-\bar \Lambda} \Big| _p + 
d\tilde\sigma ^{\Lambda-\bar \Lambda} \Big| _n &=& 
\frac{5}{9}\,[u_v(x)+d_v(x)] \, D_u^{\Lambda - \bar \Lambda}(z)
\nonumber \\
&+&\frac{2}{9}\,[s(x)-\bar s(x)] \, D_s^{\Lambda-\bar \Lambda}(z)\,.
\label{s-}
\eea
Since we now know $D_u^{\Lambda \pm \bar\Lambda}$, (\ref{s+}) and 
(\ref{s-}) allow a determination of the products 
\be
S_1(x,z) \equiv [s(x) + \bar s(x)] \, D_s^{\Lambda+\bar \Lambda}(z)
\label{S1}
\ee
and 
\be
S_2(x,z) \equiv [s(x) - \bar s(x)] \, D_s^{\Lambda-\bar \Lambda}(z)\,.
\label{S2}
\ee
In usual DIS it is the combination $(s+\bar s)$ that appears, so if 
this is taken as reasonably well determined we can extract information 
on  $D_s^{\Lambda+\bar \Lambda}(z)$ from Eq. (\ref{S1}). Of more 
interest is the question of whether the nucleon possesses intrinsic 
strange quarks, such that $s(x) \neq \bar s(x)$ \cite{z}.
Since $D_s^{\Lambda-\bar \Lambda}(z)$ is likely to be relatively large, 
a measurement of (\ref{S2}) should enable one to say whether $(s-\bar 
s)$ is compatible with zero. For further discussion of the evaluation 
of  $(s-\bar s)$ see Ref. \cite{x}.

\subsection{Cross section for unpolarized lepton and polarized nucleon 
target}

Analogously to (\ref{d-}-\ref{s-}) we now have from (\ref{b'}) 
\bea
\Delta d \tilde \sigma ^{\Lambda+\bar\Lambda} \Big| _p - 
\Delta d \tilde \sigma ^{\Lambda+\bar\Lambda} \Big| _n &=&
\frac{1}{3} [\Delta u(x) + \Delta \bar u(x) - \Delta d(x)- \Delta \bar 
d(x) ]\,
D_u^{\Lambda + \bar \Lambda}(z)\,,
\label{Deltad+}
\\
\Delta d \tilde \sigma ^{\Lambda-\bar\Lambda} \Big| _p - 
\Delta d \tilde \sigma ^{\Lambda-\bar\Lambda} \Big| _n &=&
\frac{1}{3} [\Delta u_v(x)  - \Delta d_v(x) ]\,
D_u^{\Lambda - \bar \Lambda}(z)\,,
\label{Deltad-}
\eea
where $\Delta q_v$ is defined as $\Delta q - \Delta \bar q$.
As stressed in \cite{x}, (\ref{Deltad+}) and (\ref{d-}) provide a 
stringent test for the reliability of the LO treatment. By taking their 
ratio one obtains in LO 
\be
\Delta A ^{\Lambda+\bar\Lambda} _{p-n} (x,z,Q^2) \equiv 
\frac{\Delta d \tilde \sigma ^{\Lambda+\bar\Lambda} \Big| _p - 
      \Delta d \tilde \sigma ^{\Lambda+\bar\Lambda} \Big| _n}
{d \tilde \sigma ^{\Lambda+\bar\Lambda} \Big| _p - 
 d \tilde \sigma ^{\Lambda+\bar\Lambda} \Big| _n} =
\frac{(g_1^p - g_1^n)_{LO}}{(F_1^p - F_1^n)_{LO}}(x,Q^2)\,,
\label{DeltaA}
\ee
where $g_1$ and $F_1$ are the usual polarized and unpolarized DIS 
structure functions, here evaluated in LO. The crucial feature of 
(\ref{DeltaA}) is that, in principle, the LHS depends on three 
variables $(x,z,Q^2)$, and only in LO should it be independent of $z$, 
the so called {\it passive variable}~\cite{x}.
It is essential to test this feature in order to have any confidence in 
the LO treatment.

For the experimental situation under discussion in this subsection we 
can write equations analogous to (\ref{s+}) and (\ref{s-}) via the 
substitutions
\be
d\tilde \sigma \to \Delta d \tilde \sigma \hspace{1cm} {\rm and} 
\hspace{1cm}
q(x) \to \Delta q(x)\,.
\ee
In this case we learn about the products
\be
\Delta S_1(x,z) \equiv [\Delta s(x) + \Delta \bar s(x)]  \,
D_s^{\Lambda+\bar \Lambda}(z)
\label{DeltaS1}
\ee
and 
\be
\Delta S_2(x,z) \equiv [\Delta s(x) - \Delta \bar s(x)]  \,
D_s^{\Lambda-\bar \Lambda}(z)\,.
\label{DeltaS2}
\ee
 Assuming $D_s^{\Lambda+\bar \Lambda}$ has been determined as in 
Section A, 
(\ref{DeltaS1}) would give valuable information about 
$(\Delta s + \Delta \bar s)$ which is only poorly determined from 
polarized 
DIS \cite{y}. And (\ref{DeltaS2}) could provide an answer to the 
intriguing question 
as to whether or not $\Delta s(x) = \Delta \bar s(x)$, see comments 
after
Eq. (\ref{S2}).

In addition the ratios
\be
\frac{\Delta S_1(x,z)}{ S_1(x,z)} = \frac{\Delta s(x) + \Delta \bar 
s(x)}
{s(x) + \bar s(x)} \,,
\label{DeltaS1/S1}
\ee
\be
\frac{\Delta S_2(x,z)}{ S_2(x,z)} = \frac{\Delta s(x) - \Delta \bar 
s(x)}
{s(x) - \bar s(x)}\,,
\label{DeltaS2/S2}
\ee
should be independent of the passive variable $z$. 

\subsection{Polarized \mbox{\boldmath $\Lambda$}
and \mbox{\boldmath $\bar\Lambda$} production with 
polarized lepton and unpolarized nucleon}

With a polarized lepton beam and unpolarized nucleon target, the 
difference between cross sections to produce $\Lambda$'s or  $\bar 
\Lambda$'s with helicity $\pm 1/2$ is given by (\ref{c'}).

To simplify the notation, let us write 
\be
d \tilde \sigma _{+0} ^{\Lambda_+ - \Lambda _-} \equiv 
d \tilde \sigma _{+0} ^{\Delta \Lambda } \,, \hspace{1cm} {\rm etc ...}
\ee 
Then, from (\ref{c'}) we obtain four equations analogous to (\ref{d-}), 
(\ref{d+}),  (\ref{s+}), (\ref{s-}):
\bea
d \tilde \sigma _{+0} ^{\Delta \Lambda + \Delta \bar \Lambda } \Big| _p 
-
d \tilde \sigma _{+0} ^{\Delta \Lambda + \Delta \bar \Lambda } \Big| _n 
&=&
\frac{1}{3} [ u(x) + \bar u(x) - d(x) - \bar d(x) ]\,
\Delta D_u^{\Lambda + \bar \Lambda}(z)\,,
\label{ddelta+}
\\
d \tilde \sigma _{+0} ^{\Delta \Lambda - \Delta \bar \Lambda } \Big| _p 
-
d \tilde \sigma _{+0} ^{\Delta \Lambda - \Delta \bar \Lambda } \Big| _n 
&=&
\frac{1}{3} [ u_v(x) - d_v(x)]\,
\Delta D_u^{\Lambda - \bar \Lambda}(z)\,,
\label{ddelta-}
\eea 
from which we can determine $\Delta D_u^{\Lambda}(z)$ and 
$\Delta D_u^{\bar \Lambda}(z)$, and 
\bea
d \tilde \sigma _{+0} ^{\Delta \Lambda + \Delta \bar \Lambda } \Big| _p 
+
d \tilde \sigma _{+0} ^{\Delta \Lambda + \Delta \bar \Lambda } \Big| _n 
&=&
\frac{5}{9} [ u(x) + \bar u(x) + d(x) + \bar d(x) ]\,
\Delta D_u^{\Lambda + \bar \Lambda}(z) \nonumber \\ &+&
\frac{2}{9} [ s(x) + \bar s(x)] \, \Delta D_s^{\Lambda + \bar 
\Lambda}(z)\,,
\label{sdelta+}
\\
d \tilde \sigma _{+0} ^{\Delta \Lambda - \Delta \bar \Lambda } \Big| _p 
+
d \tilde \sigma _{+0} ^{\Delta \Lambda - \Delta \bar \Lambda } \Big| _n 
&=&
\frac{5}{9} [ u_v(x) + d_v(x) ]
\,\Delta D_u^{\Lambda - \bar \Lambda}(z) \nonumber \\ &+&
\frac{2}{9} [ s(x) - \bar s(x)] \, \Delta D_s^{\Lambda - \bar 
\Lambda}(z)\,,
\label{sdelta-}
\eea
yielding information on the products 
\be
 S_3(x,z) \equiv [s(x) + \bar s(x)]\,\Delta D_s^{\Lambda+\bar 
\Lambda}(z)\,,
\label{S3}
\ee
\be
S_4(x,z) \equiv [s(x) - \bar s(x)]\,\Delta D_s^{\Lambda-\bar 
\Lambda}(z)\,.
\label{S4}
\ee
Eq.~(\ref{S3}) yields information on $\Delta D_s^{\Lambda+\bar 
\Lambda}(z)$ and (\ref{S4}) provides a further test of whether 
$s(x)=\bar s(x)$.

\subsection{Polarized \mbox{\boldmath $\Lambda$}
and \mbox{\boldmath $\bar\Lambda$} production with 
polarized nucleon and unpolarized lepton}

Analogous to (\ref{ddelta+}) and (\ref{ddelta-}), we have
\bea
d \tilde \sigma _{0+} ^{\Delta \Lambda + \Delta \bar \Lambda } \Big| _p 
-
d \tilde \sigma _{0+} ^{\Delta \Lambda + \Delta \bar \Lambda } \Big| _n 
&=&
\frac{1}{3} [\Delta u(x) + \Delta \bar u(x) - \Delta d(x) - \Delta \bar 
d(x) ]
\,\Delta D_u^{\Lambda + \bar \Lambda}(z)\,,
\label{ddelta+'}
\\
d \tilde \sigma _{0+} ^{\Delta \Lambda - \Delta \bar \Lambda } \Big| _p 
-
d \tilde \sigma _{0+} ^{\Delta \Lambda - \Delta \bar \Lambda } \Big| _n 
&=&
\frac{1}{3} [\Delta u_v(x) - \Delta d_v(x) ]
\,\Delta D_u^{\Lambda - \bar \Lambda}(z)\,.
\label{ddelta-'}
\eea 
The ratio of (\ref{ddelta+'}) and (\ref{ddelta+}) provides a further 
test of the reliability of a LO treatment. One has
\bea
\frac{
\, d \tilde \sigma _{0+} ^{\Delta \Lambda + \Delta \bar \Lambda } \Big| 
_p \,-\,
 \,d \tilde \sigma _{0+} ^{\Delta \Lambda + \Delta \bar \Lambda } \Big| 
_n\,}
{\,d \tilde \sigma _{+0} ^{\Delta \Lambda + \Delta \bar \Lambda } \Big| 
_p \,-\,
 \,d \tilde \sigma _{+0} ^{\Delta \Lambda + \Delta \bar \Lambda } \Big| 
_n\,} &=& 
{\rm function \; of} \; (x,z,Q^2) \; {\rm in \; principle} = 
\nonumber \\  &=&
\frac{(g_1^p - g_1^n)_{LO}}{(F_1^p - F_1^n)_{LO}}(x,Q^2)\,,
\eea
in LO. 
The analogues of (\ref{sdelta+}) and (\ref{sdelta-}) are obtained by 
the substitution
\be
d\tilde \sigma _{+0} ^{\Delta \Lambda \pm \Delta \bar \Lambda} \to 
d\tilde \sigma _{0+} ^{\Delta \Lambda \pm \Delta \bar \Lambda}
 \hspace{1cm} {\rm and} \hspace{1cm}
q(x) \to \Delta q(x)\,,
\ee
and yield information on
\be
\Delta S_3(x,z) = 
[\Delta s(x) + \Delta \bar s(x)]\,\Delta D _s ^{\Lambda+\bar \Lambda} 
(z)\,,
\ee
\be
\Delta S_4(x,z) = 
[\Delta s(x) - \Delta \bar s(x)]\,\Delta D _s ^{\Lambda-\bar \Lambda} 
(z)\,.
\ee
The ratios
\be
\frac{\Delta S_3(x,z)}{S_3(x,z)} = 
\frac{\Delta s(x) + \Delta \bar s(x)}{s(x) + \bar s(x)}  
\label{DeltaS3}
\ee
and 
\be
\frac{\Delta S_4(x,z)}{S_4(x,z)} = 
\frac{\Delta s(x) - \Delta \bar s(x)}{s(x) - \bar s(x)}  
\label{DeltaS4}
\ee
should be independent of $z$ in LO and should equal the ratios 
$\Delta S_1/S_1$ and $\Delta S_2/S_2$ respectively, determined in 
Section B 
[see (\ref{DeltaS1/S1}) and (\ref{DeltaS2/S2})].

\section{Conclusions}

The study of the angular distribution of the $\Lambda \to p \pi$ decay 
allows 
a simple and direct measurement of the components of the $\Lambda$ 
polarization vector. For $\Lambda$'s produced in the current 
fragmentation 
region in DIS processes, the components of the polarization vector are 
related to spin properties of the quark inside the nucleon, to spin 
properties of the quark hadronization, and to spin dynamics of the 
elementary interactions. All this information, concerning quark 
distribution functions, quark fragmentation functions and spin 
properties of elementary dynamics are essentially factorized in LO QCD 
and separated as depending on three different variables, respectively 
$x$, $z$, $y$.
The $Q^2$-evolution and dependence of distribution and fragmentation 
functions somewhat mix the three variables, but smoothly, keeping 
separated the main properties of each of the different aspects of the 
process. Moreover, such $Q^2$-dependence is perturbatively well known 
and under control.
We have discussed all different longitudinal polarization states of 
spin-$1/2$ baryons, obtainable in the fragmentation of a quark in DIS
with longitudinally polarized initial leptons and nucleons.

We have shown how one can extract new information, not attainable in 
unpolarized inclusive DIS, about parton densities, and new information, 
not extractable from $e^+e^- \to {\rm hadrons}$, about fragmentation 
functions. In particular one can learn about the poorly known
polarized strange quark density $\Delta s(x)+\Delta\bar s(x)$
and one can get some information relevant to the question
as to whether $s(x)=\bar s(x)$ and whether $\Delta s(x)=\Delta\bar s(x)$.

The connection between the theoretical quantities and the combinations 
of measured cross sections is very simple, but the challenge will be an 
experimental one, namely to have sufficient control over systematic 
errors so as to permit the combining of different measurements. We hope 
this will soon be possible.  
%
%

\acknowledgments

We are very grateful to P.~Mulders and the Theory Division of the
Physics Department, Vrije Universiteit Amsterdam,
where this work was initiated, for the warm hospitality.
M.B. is most grateful for support from the EU-TMR Program,
Contract No. CT98-0169.
U.D. and F.M. are grateful to COFINANZIAMENTO MURST-PRIN for partial support.
E.L. is grateful to the Foundation for Fundamental
Research on Matter (FOM) and the Dutch Organization for Scientific
Research (NWO) for support, and to The Leverhulme Trust for an
Emeritus Fellowship.

\begin{thebibliography}{99}
\vspace*{8pt}
\bibitem{abm00} M. Anselmino, M. Boglione, F. Murgia, \PL {B481} (2000) 
253.
\bibitem{jaf}
R.L. Jaffe, \PR{D54} (1996) 6581.
\bibitem{ekk}
J. Ellis, D. Kharzeev, A. Kotzinian, \ZP{C69} (1996) 467.
\bibitem{deflo2}
D. de Florian, M. Stratmann, W. Vogelsang, \PR{D57} (1998) 5811.
\bibitem{kbv}
A. Kotzinian, A. Bravar, D. von Harrach,
{\it Eur. Phys. J.} {\bf C2} (1998) 329.
\bibitem{kot}
A. Kotzinian, talk at the SPIN-97 VII Workshop on {\it High Energy
Spin Physics}, 7-12 July 1997, Dubna, Russia;
e-Print Archive: hep-ph/9709259.
\bibitem{bel}
S.L. Belostotski, \NP{B79} (Proc. Suppl.) (1999) 526.
\bibitem{blt}
C. Boros, J.T. Londergan, A.W. Thomas, \PR{D61} (2000) 014007.
\bibitem{al}
D. Ashery, H.J. Lipkin, \PL{B469} (1999) 263.
\bibitem{mssy}
B.-Q. Ma, I. Schmidt, J. Soffer, J-Y. Yang,
{\it Eur. Phys. J.} {\bf C16} (2000) 657.
\bibitem{y} E. Leader, A. Sidorov, D. Stamenov, \PL {B462} (1999) 189 
and references therein; C.~Bourrely, F. Buccella, O. Pisanti, P. 
Santorelli, J. Soffer, {\it Prog. Theor. Phys.} {\bf 99} (1998) 1017; G. 
Altarelli, R.D. Ball, S. Forte, G. Ridolfi, \NP {B496} (1997) 337; {\it
Acta Phys. Polon.} {\bf B29} (1998) 1145.
\bibitem{deflo1} D. de Florian, M. Stratmann, W. Vogelsang, \PRL {81}
(1998) 530; C.~Boros, J.T.~Londergan, A.W.~Thomas,
\PR{D62} (2000) 014021.
\bibitem{x} E. Christova, E. Leader, hep-ph/0007303, to appear in 
{\it Nucl. Phys.} {\bf B}.
\bibitem{z} See e.g. X. Song, \PR {D57} (1998) 4114; B.-Q. Ma, S.J. 
Brodsky, \PL {B381} (1996) 317.
\end{thebibliography}
\end{document}